%% file: paper.tex
\documentclass[5p]{elsarticle}

\usepackage{graphicx}
\usepackage{amsfonts}
\usepackage{amsmath}
\usepackage{verbatim}
\usepackage{color}

\newcommand{\half}{\frac{1}{2}}
\newcommand{\nn}{\nonumber}
\newcommand{\del}{\partial}
\DeclareMathOperator{\Tr}{Tr}

\DeclareMathOperator{\erfc}{erfc}
\newcommand{\eqn}[1]{ \begin{equation} #1 \end{equation} }

\begin{document}

\title{
\vspace{-24pt}
\begin{flushright}
{\footnotesize TCDMATH-11-09, ADP-11-23/T745}
\end{flushright}
\vspace{12pt}
Polynomial Filtered HMC -- an algorithm for lattice QCD with dynamical
  quarks} 
\author[cssm,tcd]{Waseem Kamleh}
\address[cssm]{Special Research Centre for the Subatomic Structure of Matter and Department of Physics, University of Adelaide 5005, Australia. }
\author[tcd]{Mike Peardon}
\address[tcd]{School of Mathematics, Trinity College, Dublin 2, Ireland. }


\begin{abstract}
Polynomial approximations to the inverse of the fermion matrix are used to filter the dynamics of the upper energy scales in HMC simulations. The use of a multiple time-scale integration scheme allows the filtered pseudofermions to be evolved using a coarse step size. We introduce a novel generalisation of the nested leapfrog which allows for far greater flexibility in the choice of time scales. We observe a reduction in the computational expense of the molecular dynamics integration of between 3--5 which improves as the quark mass decreases.
\end{abstract}

\maketitle

\section{Introduction \label{sec:intro}}
\input{intro}
\section{Polynomial Filtered Hybrid Monte Carlo Algorithm \label{sec:theory}}
\input{theory}

\section{Simulation Results \label{sec:results}}
\input{results}

\section{Conclusions \label{sec:conclude}}
\input{conclude}

\section{Ackowledgements\label{sec:ack}}
\input{acknowledge}
\appendix
\section{Integrator Analysis \label{sec:analysis}}
\input{appendix}

\bibliographystyle{cpc}
\bibliography{reference}

\end{document}

%% file: intro.tex

Studying the lattice representation of quantum chromodynamics (QCD) with light
quarks remains a 
numerically intensive challenge, requiring large-scale computing resources. 
The quark fields in the path integral can not be manipulated 
easily on the computer and the most practical means of proceeding is to 
integrate them analytically and deal with the resulting fermion matrix 
determinant stochastically. After this step, a path integral over the gauge 
fields alone remains and since this is an ordinary integration problem, it 
can be tackled by the Monte Carlo method. At present, the best means of
proceeding is to use an importance sampling technique, with a Markov chain of
gauge field configurations being generated and used for subsequent stochastic
estimation. The most widely used technique is the Hybrid Monte Carlo (HMC)
algorithm \cite{hmc}. Here, a fictitious time co-ordinate is introduced along 
with momentum degrees of freedom conjugate to the gauge fields and a Hamiltonian
describing evolution in this new simulation time. 

The difficulty with manipulating quark fields haunts this technique, making 
each application of the Markov process computationally intensive. Over the past
ten years, the intensive use of Hybrid Monte Carlo in large-scale production
runs means a good deal of practical experience has been gained. This experience
has exposed the difficulties with approaching the physical quark mass and 
continuum limits of lattice QCD. Recent empirical observations 
\cite{Schaefer:2009xx} suggest the 
different topological sectors that contribute to the QCD vacuum are connected
only  weakly via the Markov transitions of most commonly used versions of HMC.
A theoretical study of the approach to the continuum limit suggests that HMC 
is a non-renormalisable algorithm \cite{Luscher:2011qa}, which implies the 
computational cost 
of generating gauge field configurations with smaller lattice spacings is 
unpredictable. 

These observations and unresolved questions strongly suggest that developing 
new ideas and 
modifications to algorithms such as HMC remains useful. There has been a great
deal of activity over the past ten years developing the toolkit for numerical
simulations, including Hasenbusch's mass preconditioning
\cite{Hasenbusch:2001ne}, RHMC \cite{Clark:2006fx}, the use of 
different integrator schemes
\cite{Kennedy:2009fe, Chin:2000zz, OmelyanForceGradient}
and domain decomposition via the Schwartz alternating 
procedure \cite{Luscher:2003vf,Luscher:2005rx}. Some of these developments 
have enabled the first studies close to the theory parameters needed to make 
contact to physical data reliably \cite{Aoki:2009ix}.  There is still 
significant activity
aiming to understand and improve HMC still further \cite{Luscher:2009eq, 
Kennedy:2009fe}.  A review of recent developments can be found in 
Ref.~\cite{Jung:2010jt}.

When using HMC, there are two obstacles to pushing down the quark mass in the
simulation. Both stem from the fact that the fermion determinant must be 
represented stochastically using pseudofermions with a non-local action that
features the inverse of the fermion matrix. The first hindrance is that the
condition number of the fermion matrix increases at lighter quark mass, 
causing a large increase in the number of conjugate gradient iterations 
required to solve the linear system.  Solving this system is required at each 
molecular dynamics step, which means the numerical expense of the inversions 
is the dominant cost for generating dynamical gauge fields. The second 
hindrance, an amplification of the first, is that the molecular dynamics 
step-size must be decreased as the quark mass is reduced in order to maintain 
control over the rapid fluctuations driven by the pseudofermion field. 

Polynomial approximations to the inverse have been used to define fermion
algorithms for some time \cite{Luscher-multiboson, Frezzotti:1997ym}. Previous 
explorations in the Schwinger model \cite{mikeandjim} showed that the 
introduction of a polynomial filter gave a cheap means of controlling the
short-time-scale fluctuations in the molecular dynamics integration of HMC by
introducing a separation of time scales in the molecular dynamics by 
directly factoring the pseudofermion action into multiple parts. In this paper, 
we describe the polynomial filtering algorithm and present an investigation of 
its behaviour when applied to Monte Carlo studies of QCD. 
The paper is
organised as follows: in Sec. \ref{sec:theory} the method is described and some
details of our implementation are discussed. Simulation results are presented in
Sec. \ref{sec:results} and conclusions are drawn in Sec. \ref{sec:conclude}

%% file: theory.tex

The most powerful method for carrying out importance sampling Monte Carlo 
to estimate integrals over a large number of degrees of freedom $q$, of 
the form 
\begin{equation}
   \langle {\cal O} \rangle = \frac{1}{Z}\int\!{\cal D}q\;\; {\cal O}(q) 
      \;e^{-S(q)},
\end{equation}
is to develop a Markov process with $\frac{1}{Z} e^{-S(q)}$ as its fixed-point
probability distribution. Here, $Z$ normalises this probability measure. 
The sequence of configurations generated by repeated applications of the 
process can then be used as an appropriate ensemble for importance-sampling 
estimation of $\langle{\cal O}\rangle$.

Hybrid Monte Carlo (HMC) is just such a Markov chain Monte Carlo technique, 
comprising of two component parts. New configurations of an extended system, 
$(p,q)$ are proposed to a Metropolis test, and are accepted stochastically 
with probability 
\begin{equation}
  {\cal P}_{\rm acc} = \min (1, e^{-\Delta {\cal H}}),
\end{equation}
based on the change in the Hamiltonian,
\begin{equation}
{\cal H}(p,q) = \frac{1}{2} \sum p^2 + S(q).
\end{equation}
The extended system doubles the degrees of freedom, adding a conjugate variable 
$p$ to each co-ordinate $q$. Since the Metropolis test ensures the probability 
of a configuration of this new system occurring in the Markov chain is given by 
$\exp\{-{\cal H}\} = \exp\{-T(p)\} \times \exp\{-S(q)\}$, the separable 
product of 
probability densities of $p$ and $q$, these two sets of variables are 
independent random numbers. Subsequently, the new variables $p$ can be 
discarded, leaving an ensemble of configurations of $q$ with the desired 
probability distribution. 

To ensure a useful acceptance probability for the Metropolis test, ${\cal H}$
is interpreted as the Hamiltonian describing the dynamics in the phase space
generated by $(p,q)$, where $p$ is treated as a canonical momentum conjugate 
to $q$. In practical simulations, it is the molecular dynamics that requires
most of the computer time for the Markov chain to evolve. 

This construction strongly suggests that improving the efficiency of the 
algorithm requires both accelerating and improving the accuracy of the 
molecular dynamics integration process.
As with many complex systems, the classical dynamics of the degrees of freedom
of lattice QCD incorporates interactions with a broad range of characteristic
time scales. Any attempt to improve the performance of molecular dynamics
will be aided by describing a method that allows the separate dynamical 
scales to be treated differently. As a starting point, consider the 
multiple time-scale scheme of Sexton and Weingarten \cite{sexton-weingarten}. 
In this construction, 
an integrator which captures the different dynamical scales of different 
parts of the action can be defined. 
Integrators to evolve the system in the new time co-ordinate are constructed 
from two basic time-evolution operators, generated by kinetic and potential 
energy terms. Their effect on the system co-ordinates, $(p,q)$ is 
\begin{equation}
   V_T(h): \{p,q\} \longrightarrow \{p, q+h\; p\}, 
\end{equation}
and
\begin{equation}
   V_S(h): \{p,q\} \longrightarrow \{p - h\;\partial S, q\},
\end{equation}
where $\partial S$ is the ``extended force'' due to the action $S.$
The simplest time-reversible integrator is then built from the leap-frog 
scheme, 
\begin{equation}
   V(h) = V_S(\frac{h}{2}) V_T(h) V_S(\frac{h}{2}),
\end{equation}
and repeated applications of this building block evolves the system with
the Hamiltonian conserved up to corrections of ${\cal O}(h^2)$.

\subsection{Multiple time-scales in molecular dynamics integrators}

If the action and thus the Hamiltonian is split into two parts $H_1$ and 
$H_2$,
\begin{equation}
  {\cal H} = \underbrace{\frac{1}{2}\sum p^2 + S_1(q)}_{H_1} + 
             \underbrace{S_2(q)}_{H_2},
\end{equation}
where $S_1$ is an action capturing the high-frequency modes and $S_2$ captures 
the low-frequency ones, then a generalised leap-frog integrator can be written.
The two leap-frog integrators for the two Hamiltonians are defined as
\begin{equation}
  V_1(h) = V_{S_1}(\frac{h}{2}) V_T(h) V_{S_1}(\frac{h}{2}), 
\end{equation}
\begin{equation}
  V_2(h) = V_{S_2}(h), 
\end{equation}
and a compound integrator for the full Hamiltonian can be constructed by 
combining the two components:
\begin{equation}
  V(h) = V_2(\frac{h}{2}) \left[ V_1(\frac{h}{m})\right]^m V_2(\frac{h}{2}), 
\end{equation}
where $m \in \mathbb{N}$. This compound integrator effectively introduces two 
time-scales into the evolution, $h$ and $h/m$. 

Since the force term for $S_1$ must be evaluated many more times than that for
$S_2$, the method will only be useful if two conditions are met 
simultaneously. First, as already suggested, the two actions must effectively 
split the dynamical scales into high and low-frequency parts in $S_1$ and $S_2$,
and also, the evaluation of $\partial S_1$ must be computationally much cheaper 
than that for $\partial S_2$.

\subsection{HMC for Lattice QCD}

When using HMC to generate an ensemble of gauge configurations for importance
sampling Monte Carlo estimation of the path integral of lattice QCD, a few 
details arise. The first is that the integration variables are elements of 
the gauge group, $SU(3)$, and the gauge invariant Haar measure must be used. 
This can be achieved by making the conjugate momenta elements of the Lie 
algebra of the gauge group, and by modifying the action of $V_T$ slightly;
\begin{equation}
  V_T(h): (P,U) \longrightarrow (P,e^{i h P} U).
\end{equation}
Correspondingly, the effect of the evolution operator $V_S$ must then be 
\begin{equation}
  V_S(h): (P,U) \longrightarrow (P - h \Sigma, U), 
\end{equation}
where
\begin{equation}
  \Sigma = \frac{\Lambda - \Lambda^\dagger}{2i} - 
     \frac{1}{N}\mbox{ImTr}\Lambda,
\end{equation}
with
\begin{equation}
  \frac{dS}{dt} = \mbox{ReTr} \left\{\Lambda \frac{dU}{dt}\right\}.
\end{equation}

While this defines molecular dynamics on the group manifold, the action for 
QCD with the quark field dynamics included requires some manipulation. After
integrating out the Grassmann variables representing the quark fields, the 
importance sampling probability measure for two degenerate flavours of quarks 
becomes 
\begin{equation}
  {\cal P}[U] = \frac{1}{Z}\;\det\!{}^2 M[U]\;\; e^{-S_G[U]} 
         = \frac{1}{Z} e^{-S_{\rm eff}[U]}.
\end{equation}
The effective action for this measure is given by
\begin{equation}
  S_{\rm eff}[U] = -2 \mbox{Tr log} M[U]\;\; + S_G[U].
\end{equation}
Unfortunately, evaluating this action and the force terms that arise from the
molecular dynamics is very cumbersome on the computer. A solution requires the
introductions of ``pseudofermions'', which represent the fermion determinant 
as a Gaussian integral;
\begin{equation}
  \det\!{}^2 M[U] = \int\!\!{\cal D}\phi{\cal D}\phi^* \;\;
           e^{-\phi^* (M^\dagger M)^{-1} \phi},
\end{equation}
and the new effective action for the system is 
\begin{equation}
  S_{\rm eff}[U,\phi,\phi^*] = 
           \phi^* K^{-1} \phi + S_G[U],
     \label{eqn:pseudofermi}
\end{equation}
where $K = M^\dagger M.$ The $\gamma_5$-Hermiticity property of the fermion matrix has been exploited 
here; since $M^\dagger = \gamma_5 M \gamma_5,$ $\det M^\dagger  = \det M$ and 
hence $\det K = \det\!{}^2 M$.
The molecular dynamics force term for the pseudofermions can now be computed;
this requires the application of the inverse of the fermion matrix. While 
this step is computationally very expensive, it is at least tractable. 
Since the action has two components, it seems natural to use the 
Sexton-Weingarten scheme to integrate both parts with different time-scales
appropriate to their dynamics. Unfortunately, at light quark masses the 
high-frequency dynamics occurs in the pseudofermion action, which also has the 
most expensive force evaluation. 

\subsection{The polynomial filtered hybrid Monte Carlo algorithm}

The requirement that the integrator for the shortest time-step has a force
term that is computationally cheap to implement leads to using low-order
polynomial approximations to the inverse of the fermion matrix. 
Clearly, a short polynomial satisfies the condition that its force is cheap
to evaluate. The hope is that such a simple approximation might be able to 
mimic the short-distance part of the propagator, and so capture the 
high-frequency dynamics expected in this part of the system. 

Making use of the identity which holds for any polynomial, ${\cal P}$ of the 
(two flavour) fermion matrix 
\begin{equation}
  \det K = \frac{\det K {\cal P}(K) }{\det {\cal P}(K)},
    \label{eqn:detratio}
\end{equation}
suggests introducing two auxiliary integrals for each of these determinants 
separately,
\begin{equation}
  \det\!{}^2 M = \int\!\!
     {\cal D}\phi {\cal D}\phi^*
     {\cal D}\chi {\cal D}\chi^*
        \;\;\;e^{-\phi^* ({\cal P} K)^{-1} \phi 
                - \chi^* {\cal P} \chi}.
\end{equation}
The action for QCD then becomes
\begin{equation}
  S_{\rm filter}[U] = 
            \phi^* ({\cal P} K)^{-1} \phi 
          + \chi^* {\cal P}  \chi 
          + S_G[U].
\end{equation}
This action is separated into two parts, $S_1$ and $S_2$, 
\begin{equation}
  S_1 = 
            \phi^* ({\cal P} K)^{-1} \phi ,
\end{equation}
\begin{equation}
  S_2 = 
          \chi^* {\cal P}  \chi 
          + S_G[U].
\end{equation}
In the limit that a perfect representation of the inverse is constructed, {\it 
i.e.} $K {\cal P}(K) \rightarrow I$, the first action becomes independent of 
the gauge fields and so induces no molecular dynamics force. Conversely,
as the order of the polynomial reduces, ${\cal P} \rightarrow I$ and then
the system reproduces the pseudofermion action of Eqn.~(\ref{eqn:pseudofermi}).
The polynomial can thus be used to capture as much or as little of the 
dynamics of the fermion action as is necessary to extract the high-frequency
modes as cheaply as possible.

\subsection{Adding additional time-scales.}

The determinant ratio identity of Eqn.~(\ref{eqn:detratio}) can be extended to
make use of more than one polynomial
\begin{equation}
  \det K = \det K {\cal P}_2(K) \times
   \frac{1}{\det {\cal P}_1^{-1}(K) {\cal P}_2(K)} \times
   \frac{1}{\det {\cal P}_1(K)}
\end{equation}
The order of the first polynomial, $n_1$ is chosen to be much less than that 
of the second, $n_2 \gg n_1$. For this construction to be useful, the two 
polynomials should be constructed so all the roots of the first are contained 
in the set of roots of the second. In this case, ${\cal P}_2$ can be written 
as the product of two polynomials, 
\begin{equation}
   {\cal P}_2 = {\cal P}_1 {\cal Q}
\end{equation}
where the order of the new polynomial ${\cal Q}$ is $n_q = n_2 - n_1$. 
Now, the fermion determinant can be re-expressed using three integral 
representations, and the resulting action for QCD becomes 
\begin{equation}
  S_{\rm 2-poly} = 
            \phi^* ({\cal P}_2 K)^{-1} \phi 
          + \chi_2^* {\cal Q}^\dagger {\cal Q}  \chi_2
          + \chi_1^* {\cal P}_1^\dagger {\cal P}_1  \chi_1
          + S_G.
\end{equation}
Judicious choice of the two polynomials, ${\cal P}_1$ and ${\cal P}_2$ should
capture the dynamics of successively increasing time-scales, and allow a
generalised version of the Sexton-Weingarten scheme to be employed. 

\subsection{Polynomial approximation to $\frac{1}{z}$}

Naturally, in order to be able to perform Monte Carlo simulations ${\cal P}(K)$ must be a real polynomial and hence possess roots in complex conjugate pairs. 
In this work we use a Chebyshev approximation to to $1/z,$
\eqn{ P_n(z) = a_n\prod_{k=1}^n (z - z_k) \approx \frac{1}{z}. }
The roots are given by \cite{Luscher-multiboson,Borici-systematic} 
\eqn{ z_k = \mu(1-\cos\theta_k) - i\sqrt{\mu^2-\nu^2}\sin\theta_k, }
with $\theta_k = \frac{2\pi k}{n+1}$.
The normalisation is then 
\eqn{a_n = \frac{1}{\mu\prod_{k=1}^n (\mu - z_k)}.}
The polynomial possesses roots which lie on an ellipse in the complex plane. The parameters $\mu$ and $\nu$ should be chosen such that the spectrum of $K$ lies within the ellipse. Within this constraint one is free to tune the two parameters. For the choice $\mu = 1,\nu = 0$ the polynomial coincides with that given by the truncation of the Taylor series expansion of $1/z$ about $z=1.$

\subsection{A generalised multi-scale leap-frog integrator}

For a two flavour simulation with up to two polynomial terms in the Hamiltonian, we introduce the possibility of simulating at four different time scales, one for each of the terms involved,
\begin{align}
\nn &S_1 = S_G,& &S_2 = \chi_1^* {\cal P}_1 \chi_1,\\
\nn &S_3 = \chi_2^* {\cal Q} \chi_2,& &S_4 = \phi_{\rm 2f}^* ({\cal P}_2 K)^{-1}
\phi_{\rm 2f}.
\end{align}
Associate a timestep $h_i$ to each term $S_i$ and a corresponding integer 
$N_i$ such that $h_i = 1/N_i.$ $h_1$ corresponds to the step-size at 
which the gauge fields 
  are updated. 
Assuming that $N_i > N_j$ for $i < j$, the Sexton-Weingarten nested leapfrog 
algorithm then requires that $N_i | N_{i-1} \; \forall\ i > 1.$ 
The restrictions this places on the various $N_i$ may not be the most 
efficient or flexible way of performing the molecular dynamics integration. 

A generalised scheme \cite{Kamleh:2011vk} in which the only 
requirement is 
\eqn{N_i | N_1 \;\forall\ i > 1,} 
can be defined. 
In a standard leapfrog algorithm, one alternates between updates $V_T$ to the gauge field $U$ and updates $V_S$ to the conjugate momenta. Let $V_i$ denote the update to $P$ corresponding to the action $S_i.$ Now, as the guide bosons are held fixed during an integration the updates $V_i$ only depend upon the gauge field. As the updates $V_i$ are additive to $P,$ it follows that the different $V_i$ commute:
\eqn{V_i(\frac{h_i}{2}) V_j(h_j) V_i(\frac{h_i}{2})  =  V_j(h_j) V_i(h_i) = V_i(h_i) V_j(h_j).}
Define the integers
\eqn{m_i = N_1 \div N_i}
to be the ratios of the scales. In order to construct our reversible integrator we first define a map
\begin{equation}
\Theta[V;m,k \in \mathbb{N}] = \left\{ 
\begin{aligned} &V\text{ if } m | k \\ 
&I\text{(the identity) otherwise.} 
\end{aligned}\right.
\end{equation}
Let $m_T$ be the lowest common multiple of $\{ m_i \},$ and let $h_T$ be the smallest time step (in our case $h_T = h_1).$ Then our integrator is
\begin{multline}
 V(h) = \prod_i V_i(\frac{h_i}{2}) \times \\
\prod_{k=1}^{m_T-1} V_T(h_T) \Big\{ \prod_i \Theta[V_i(h_i),m_i,k]\Big\} \\
  \times V_T(h_T) \prod_i V_i(\frac{h_i}{2}),
\end{multline}
where $h = m_T h_T$ is the total timestep taken by $V.$ 
While it appears cumbersome here, the above expression is straightforwardly 
implemented in software. We demonstrate this with a pseudocode implementation 
here. Denote by $\{a \equiv b \mod m\}$ the usual notion of congruence modulo $m.$ Then we can implement the generalised integrator as follows:
\begin{itemize}
\item For each term in the action $S_i$ perform an initial half-step $V_i(\half h_i)$ updating $P.$
\item Loop over $j=1$ to $N-1$ 
\begin{itemize}
\item Apply $V_T(h)$ to update $U.$
\item If $\{0 \equiv j \mod m_i\}$ apply  $V_i(h_i)$ to update $P$
\end{itemize}
\item Apply $V_T(h)$ to update $U.$
\item For each term in the action $S_i$ perform a final half-step $V_i(\half h_i)$ updating $P.$
\end{itemize}

 The advantage of the generalised integrator is that it allows finer control 
when there are many different scales. An analysis of the finite-step size 
errors for the generalised integrator is provided in \ref{sec:analysis}.

\subsection{Polynomial force term}

For completeness, we review here the molecular dynamics force generated by the 
polynomial terms \cite{Frezzotti:1997ym,phmc}. Given a polynomial
\eqn{ {\cal P}(K) = a_n \prod_{i=1}^n (K - z_i) }
and the following action
\eqn{ S_P = \phi^\dagger {\cal P}(K) \phi,}
we define the auxiliary fields
\begin{align}
\chi_j &= \prod_{i > j} ( K - z_i) \phi, \\
\eta_j &= a_n \prod_{i < j} (K - z_i^*) \phi,
\end{align}
Then we have
\eqn{ \frac{dS_P}{dt} = \Tr\left\{\sum_{j=1}^n \left( \eta_j^\dagger \frac{dK}{dt}\chi_j \right)\right\}. }
Although the above equation appears to require ${\cal O}(n^2)$ matrix operations
one can reduce this to ${\cal O}(n)$  by trading off for storage of ${\cal
  O}(n)$ intermediate fields \cite{Frezzotti:1997ym}.


%% file: results.tex

\begin{figure*}[!t]
\centering
\includegraphics[angle=90,height=0.28\textheight,width=0.45\textwidth]{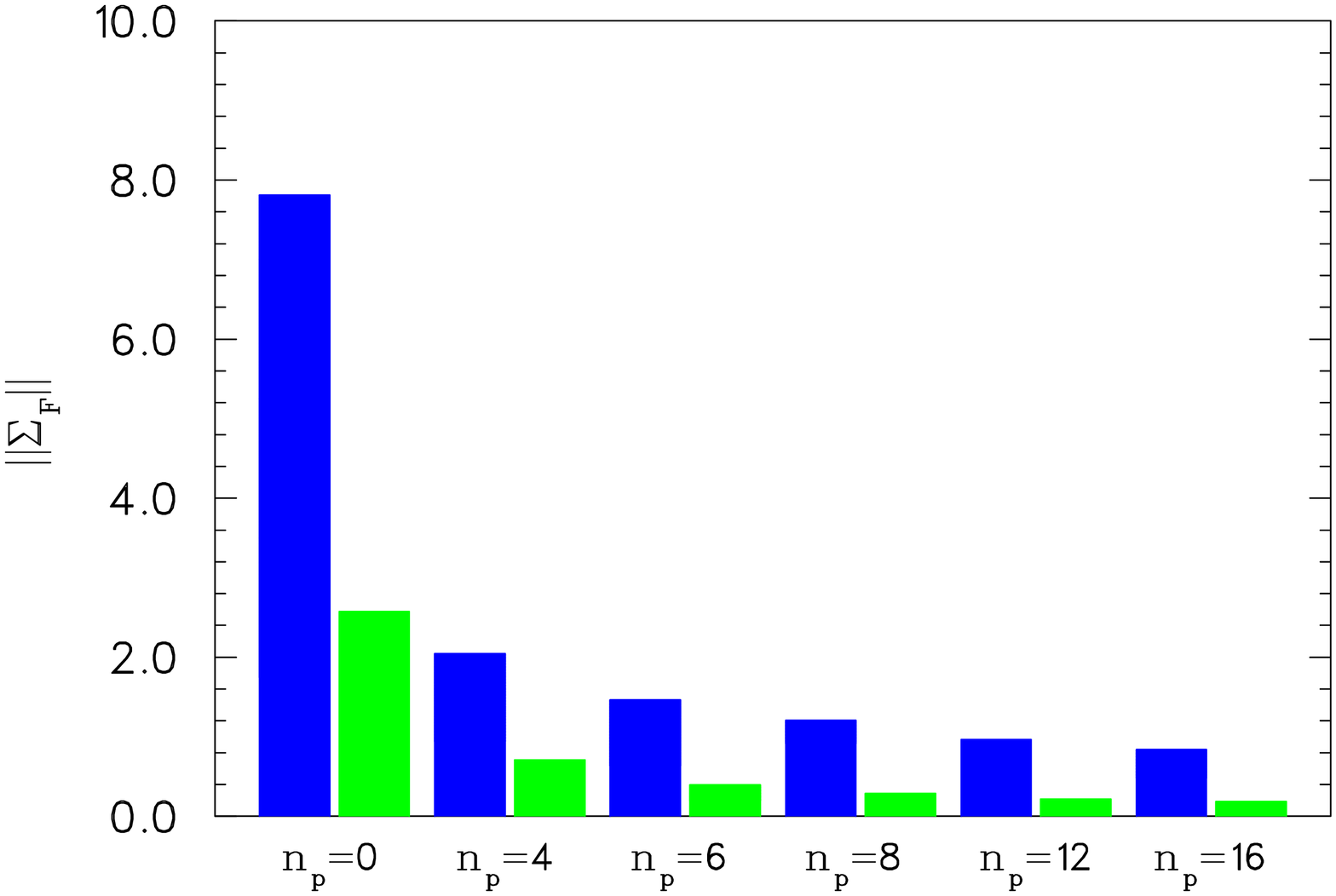}
\includegraphics[angle=90,height=0.28\textheight,width=0.45\textwidth]{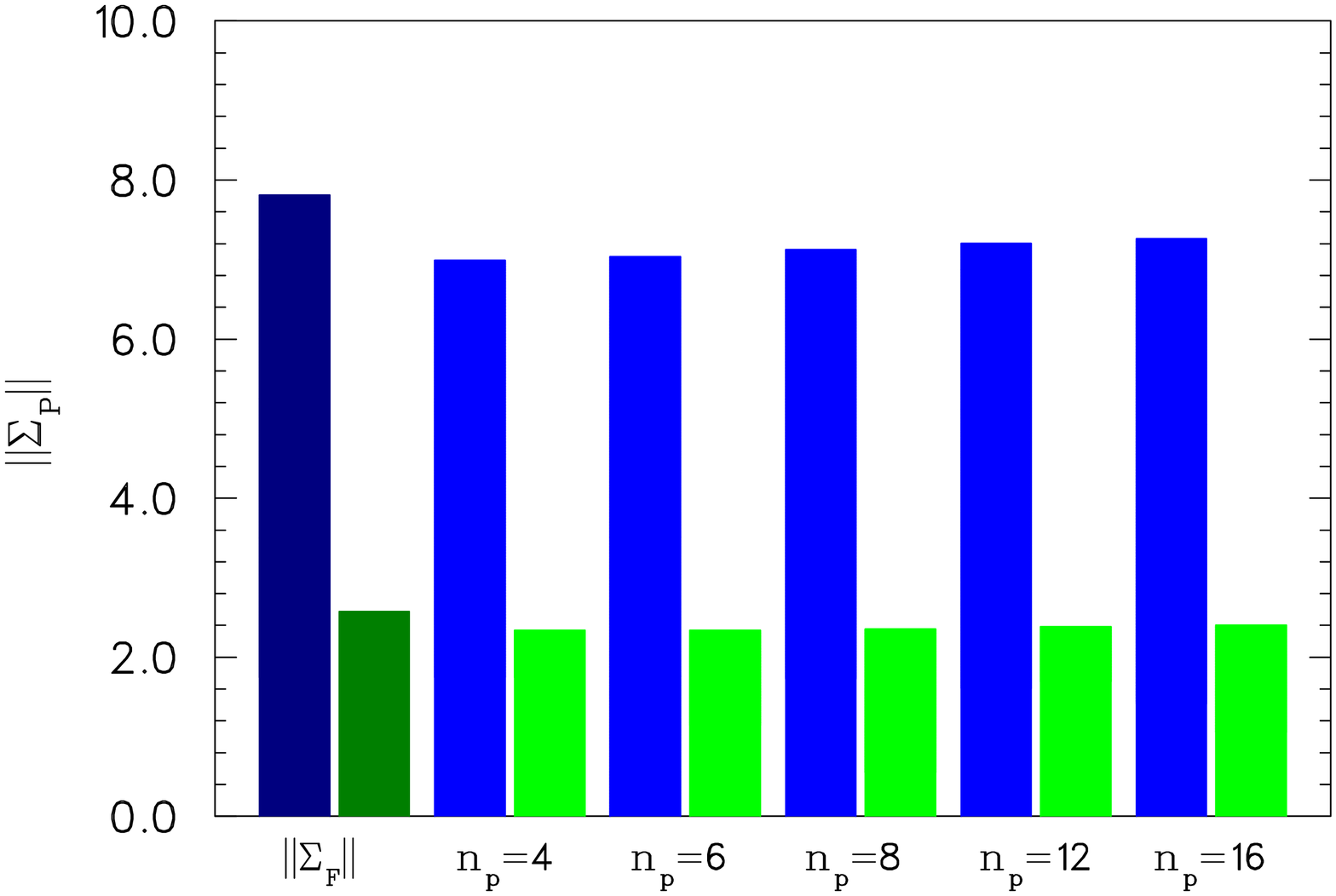}

\caption{\label{fig:force-p}The size of the force due to the pseudofermions $||\Sigma_F||$ (left) and that due to the polynomial term $||\Sigma_P||$ (right), as a function of (single) polynomial filter order $n_p.$ The maximum and average force size is shown for $\kappa=0.1575.$ For comparison, the values for $||\Sigma_F||$ with $n_p=0$ are also shown as the leftmost darker coloured bars in the graph for $||\Sigma_P||$ (right). }
\end{figure*}

\begin{figure*}[!t]
\centering
\includegraphics[angle=90,height=0.28\textheight,width=0.45\textwidth]{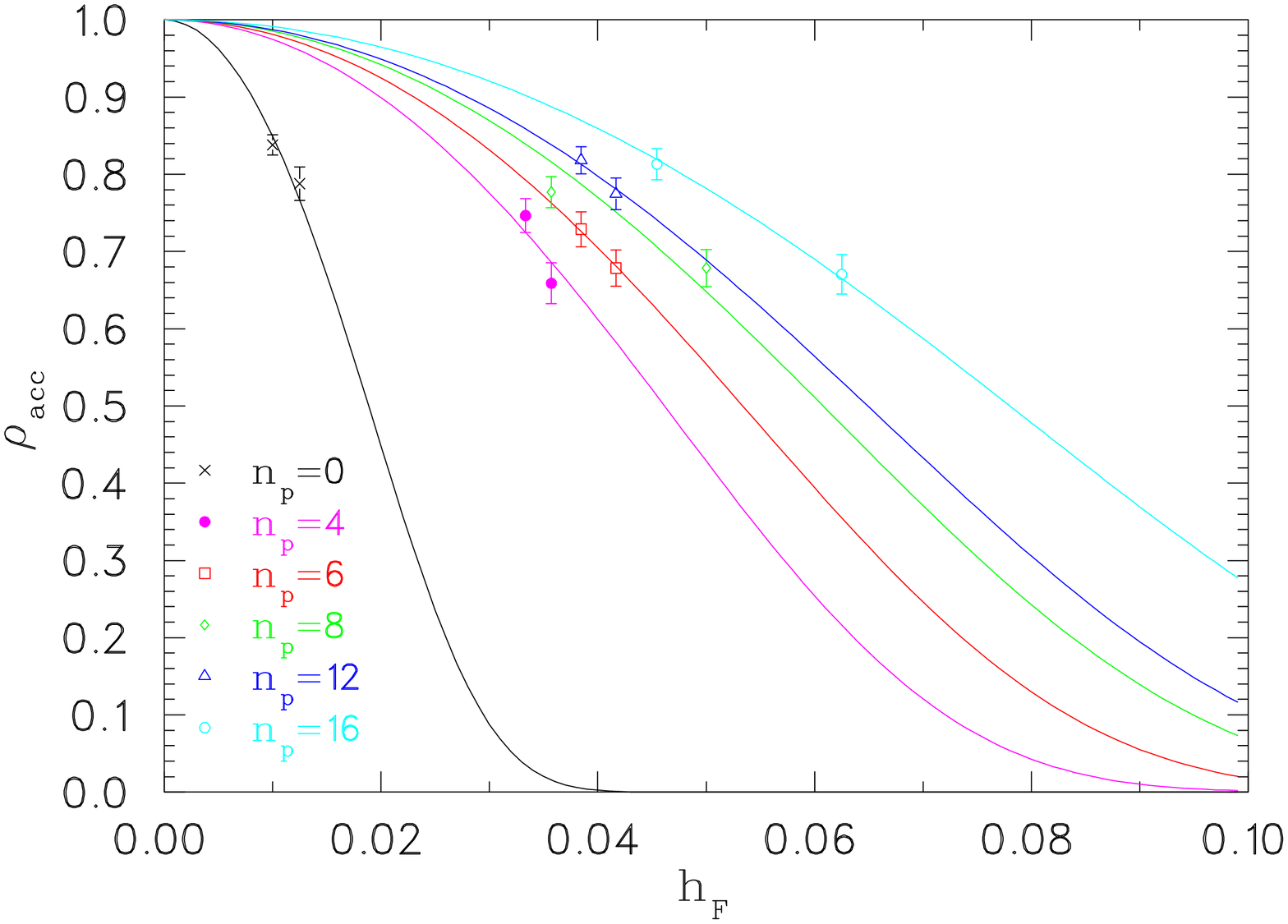}
\includegraphics[angle=90,height=0.28\textheight,width=0.45\textwidth]{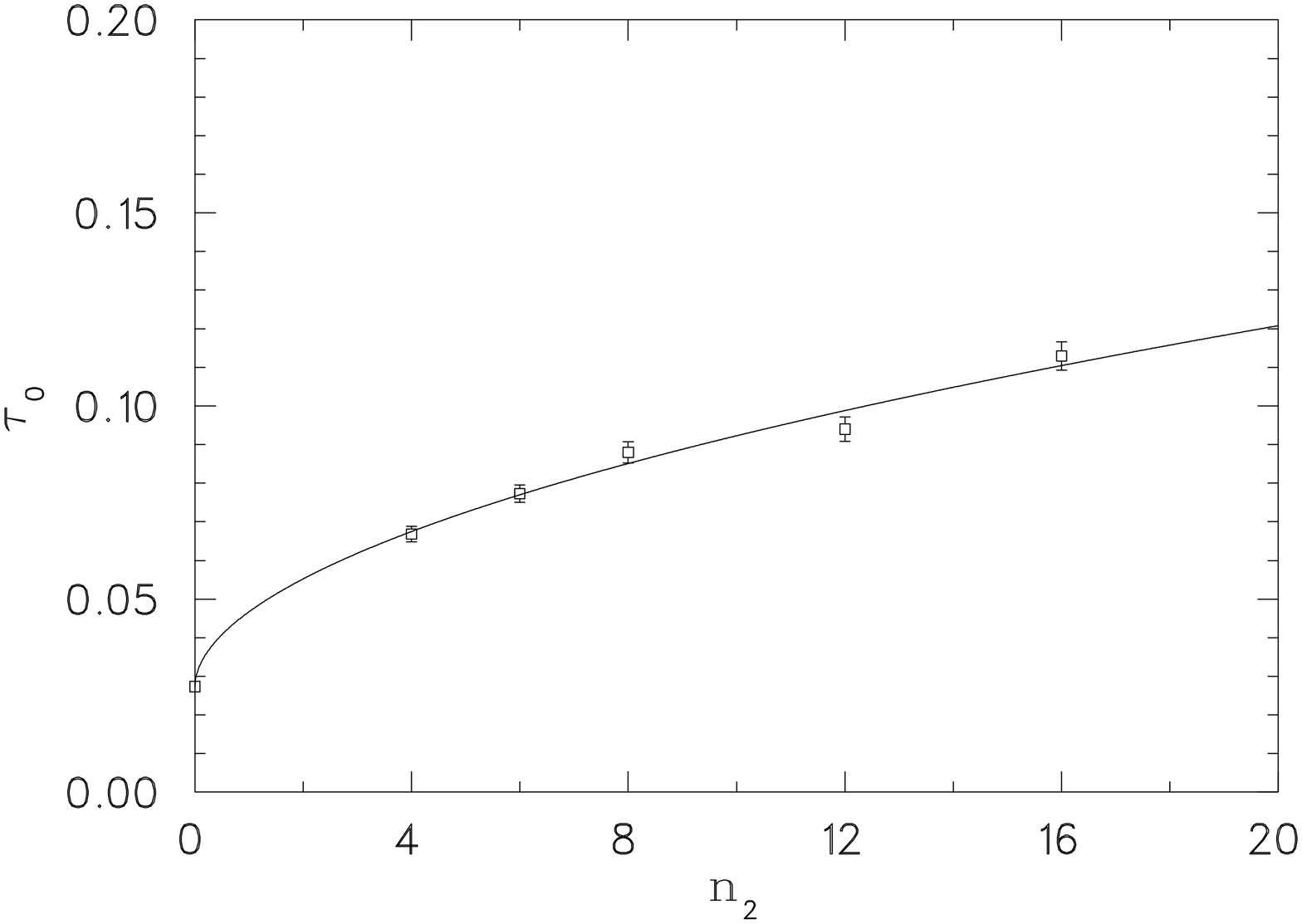}
\caption{\label{fig:accept-p}(Left) Trajectory acceptance rate as a function of pseudo-fermion step size for a single polynomial filter of various order $n_p.$ (Right) Fit of the characteristic scale $\tau_0$ as a function of polynomial filter order $n_2 = n_p.$ Results are shown for $\kappa=0.1575.$}
\end{figure*}

Results are calculated on lattices using the Wilson gauge action and the unimproved even-odd preconditioned Wilson fermion action. In order provide a straightforward comparison with alternative algorithms, we chose physical parameters that have been used elsewhere \cite{Luscher:2005rx,Urbach:2005ji}. The gauge coupling is set to $\beta=5.6.$ We work at two quark masses, $\kappa=0.1575$ and $0.15825$ which correspond to pion masses of $600$ and $400$ MeV respectively.

We begin by examining the effectiveness of a single polynomial filter $(n_q = 0).$ We measure the size of the force $\Sigma_F$ due to the filtered pseudofermion term $S_F$ and of the force $\Sigma_P$ due to the polynomial term $S_P.$
 Each element of the force is in the Lie algebra $su(3)$ hence as in other work\cite{Luscher-schwarz} we use the Lie norm $||X||^2 = -2\Tr{X^2}.$ Across a configuration we measure the mean and maximum values of $||\Sigma||.$ These values do not vary much from one trajectory to the next. 
Tuning of the parameters $\mu$ and $\nu < \mu$ is done by choosing the values which minimise a sample measurement of the force on a single configuration.

The left-hand plot in Figure~\ref{fig:force-p} shows typical mean and maximum forces due to the (filtered) pseudofermions. We see that even using very short polynomials gives significant reduction in the force, which continues to decrease as one increases the order of the polynomial $n_p.$ We note however that the rate of the reduction in the force is sub-linear in $n_p.$

The right-hand plot in Figure~\ref{fig:force-p} shows typical mean and maximum
forces due to the polynomial term. We see that this is roughly independent of
the size of the polynomial, and is approximately equal to the value for $n_p=0.$
This indicates the the size of the force is dominated by the shortest scale
present, which strongly supports the motivation for separating the time scales.

\begin{figure*}[!t]
\centering
\includegraphics[angle=90,height=0.28\textheight,width=0.45\textwidth]{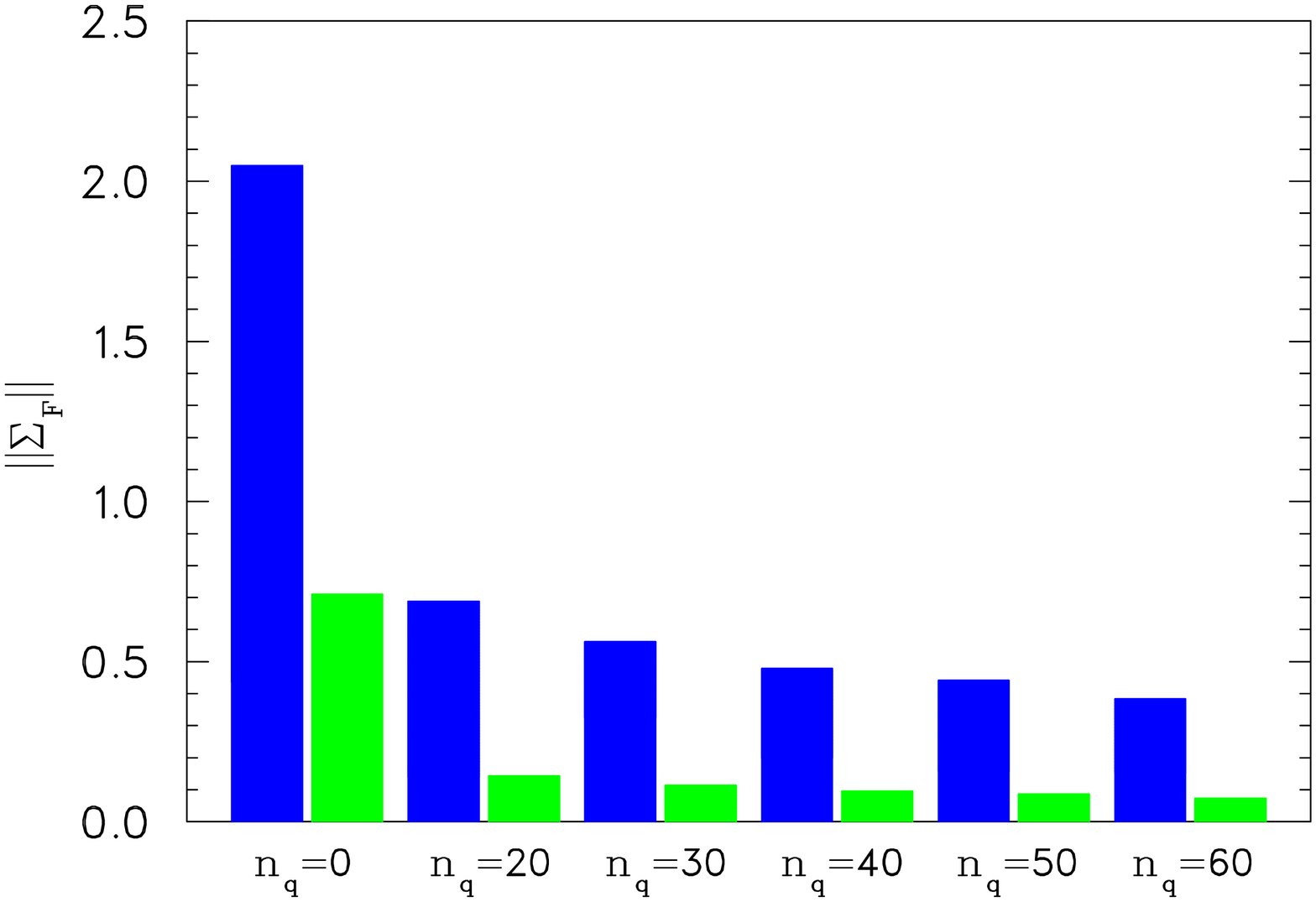}
\includegraphics[angle=90,height=0.28\textheight,width=0.45\textwidth]{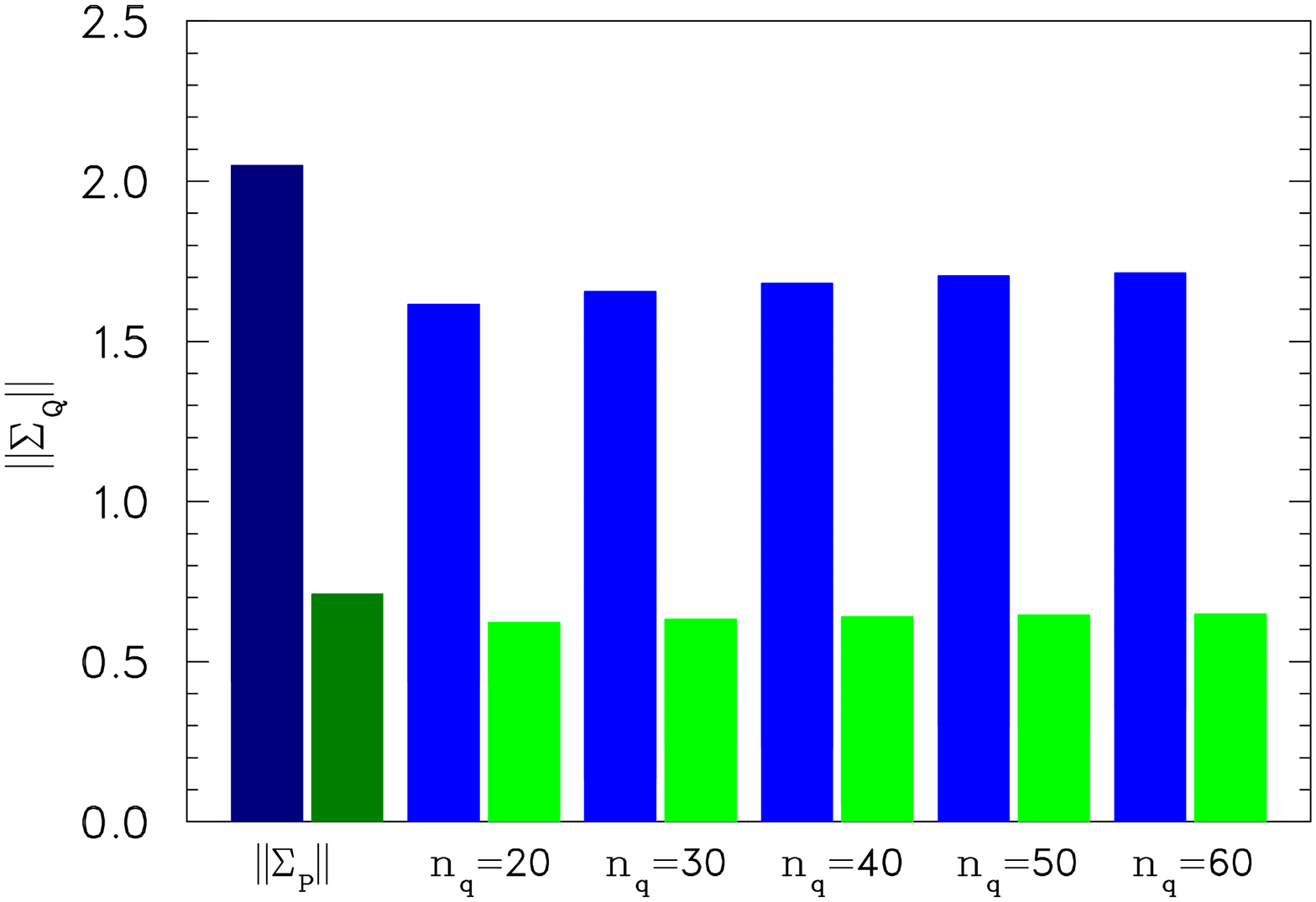}

\caption{\label{fig:force-pq}The size of the force due to the pseudofermions $||\Sigma_F||$ (left) and that due to the polynomial term $||\Sigma_Q||$ (right), as a function of polynomial filter order $n_q$ with $n_p=4$ fixed. The maximum and average force size is shown for $\kappa=0.1575.$ For comparison, the values for $||\Sigma_P||$ with $n_p=4,n_q=0$ are also shown as the leftmost darker coloured bars in the graph for $||\Sigma_Q||$ (right). }
\end{figure*}

\begin{figure*}[!t]
\centering
\includegraphics[angle=90,height=0.28\textheight,width=0.45\textwidth]{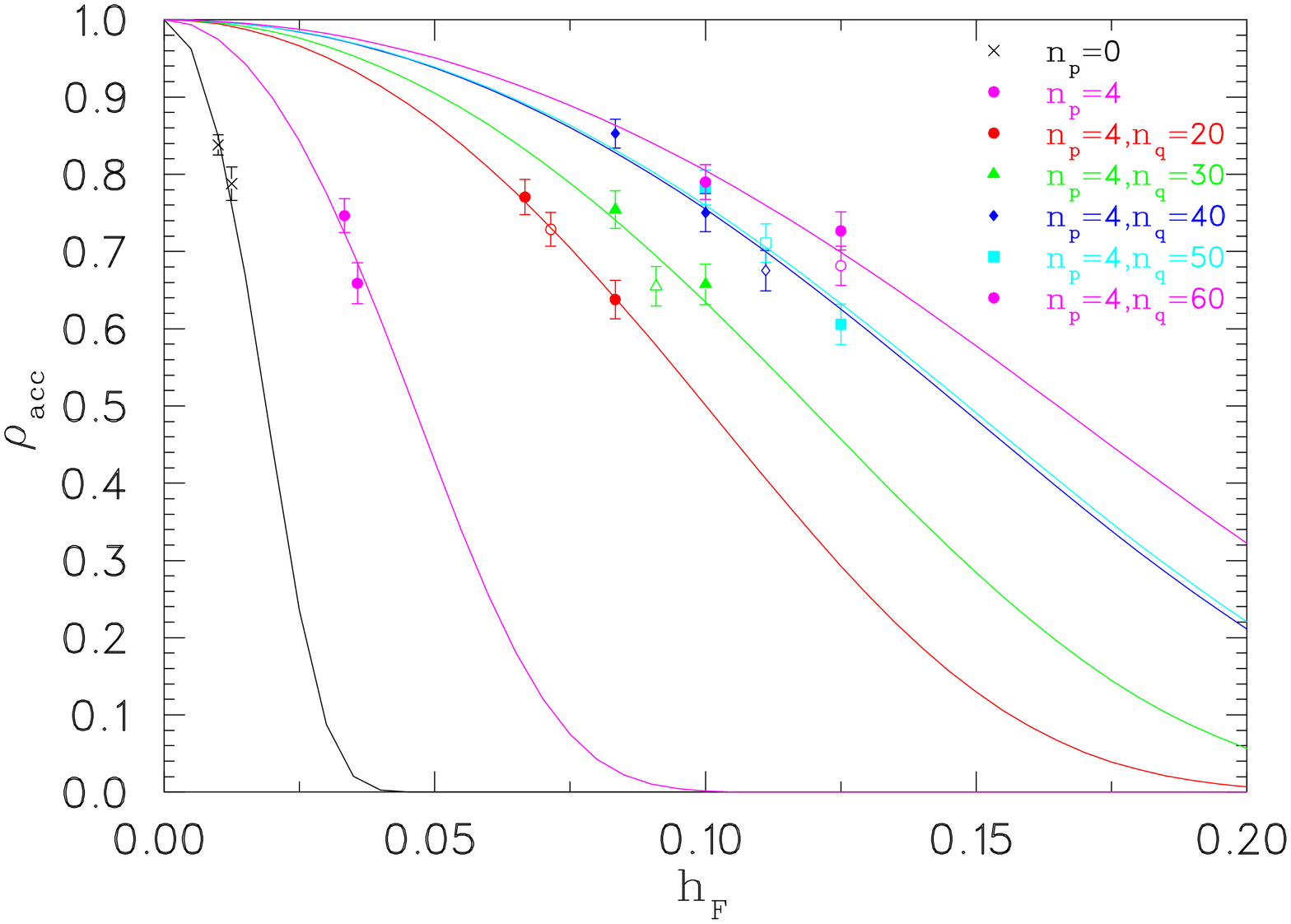}
\includegraphics[angle=90,height=0.28\textheight,width=0.45\textwidth]{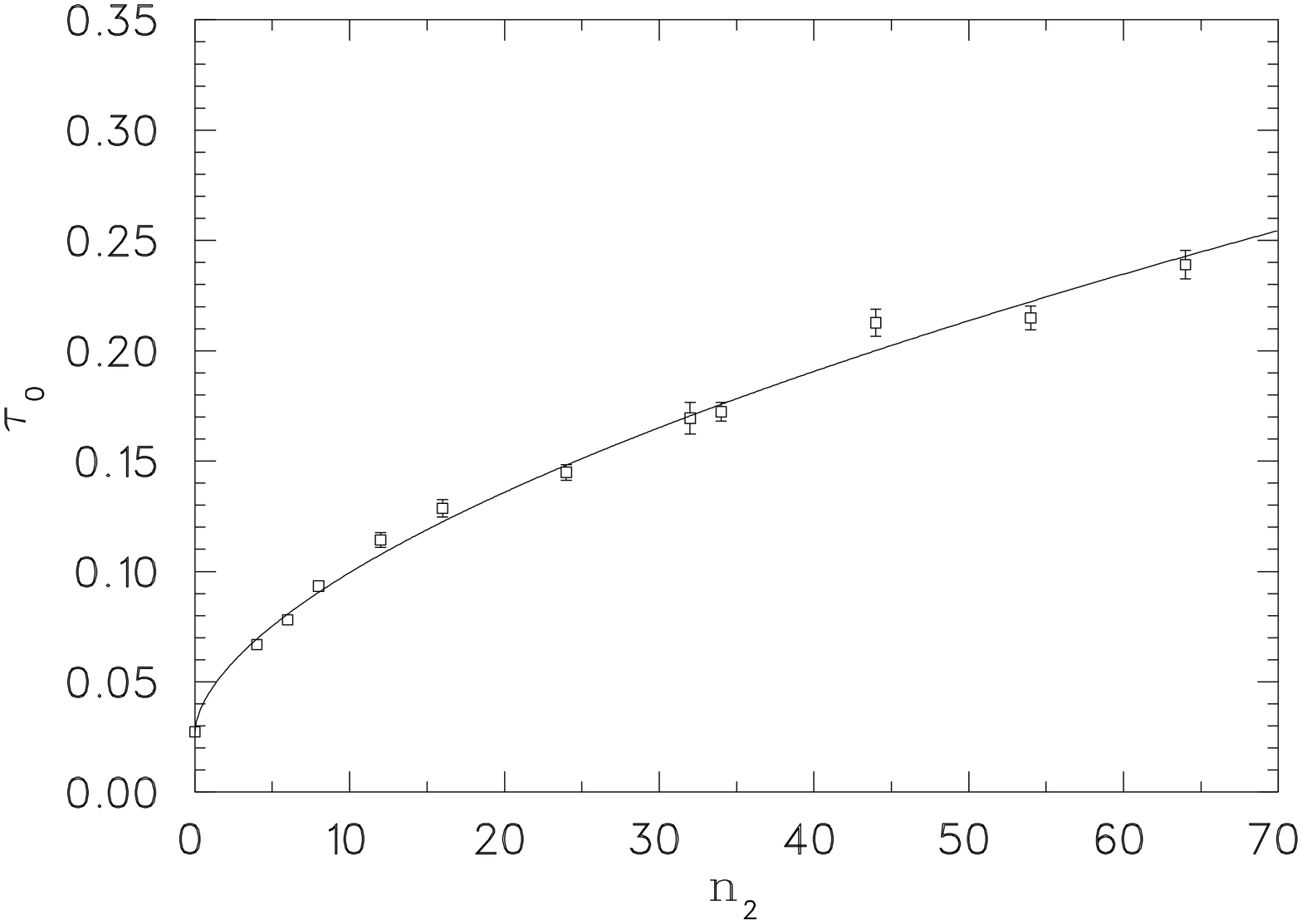}

\caption{\label{fig:accept-pq}(Left) Trajectory acceptance rate as a function of pseudo-fermion step size for a dual polynomial filter of fixed order $n_p=4$ and various order $n_q$. (Right) Fit of the characteristic scale $\tau_0$ as a function of polynomial filter order $n_2 = n_p+n_q.$ Results are shown for $\kappa=0.1575.$}
\end{figure*}

We can also measure the efficacy of the polynomial filter by considering the acceptance rate as a function of the pseudofermion step size. Figure~\ref{fig:accept-p} shows the acceptance rate as a function of $h_F$ for a single polynomial filter (of various order $n_p$). The polynomial step size was kept approximately fixed. Fits are performed using the complementary error function
\eqn{\label{eq:rho} \rho_{\rm acc} = \erfc(\frac{h_F^2}{\tau_0^2}). }
We call $\tau_0$ the characteristic scale. The characteristic scale as a function of polynomial order is shown in the right hand plot of Figure~\ref{fig:accept-p}. To gain some intuitive understanding, we can see that using a step size of half the characteristic scale will yield an acceptance rate of approximately $0.7.$ We fit the characteristic scale as function of $n_1$ as follows
\eqn{ \tau_0 = a + b n_1^c. }

The addition of a second polynomial filter does not change the qualitative behaviour of the quantities we have examined. To study the effects of adding a second polynomial, we fix $n_p=4$ and vary $n_q.$ Figure~\ref{fig:force-pq} shows the size of the force $\Sigma_F$ due to the pseudofermions and that due to the second polynomial term $\Sigma_Q.$ Even at moderate $n_q, ||\Sigma_F||$ is significantly reduced compared to its unfiltered value, by a factor of between 10-20. We see again that $||\Sigma_F||$ decreases sub-linearly with $n_q$ and that $||\Sigma_Q||$ is essentially independent of $n_q,$ being dominated by the shortest scale present.
Figure~\ref{fig:accept-pq} shows the acceptance rate as a function of $h_F$ for a 2-filter system (with $n_p=4$ and various $n_q$) on the left, and the characteristic scale $\tau_0$ as a function of $n_2 = n_p + n_q$ on the right. We again perform a fit for $\tau_0$ as before.

\begin{figure*}[!t]
\centering
\includegraphics[angle=90,height=0.28\textheight,width=0.45\textwidth]{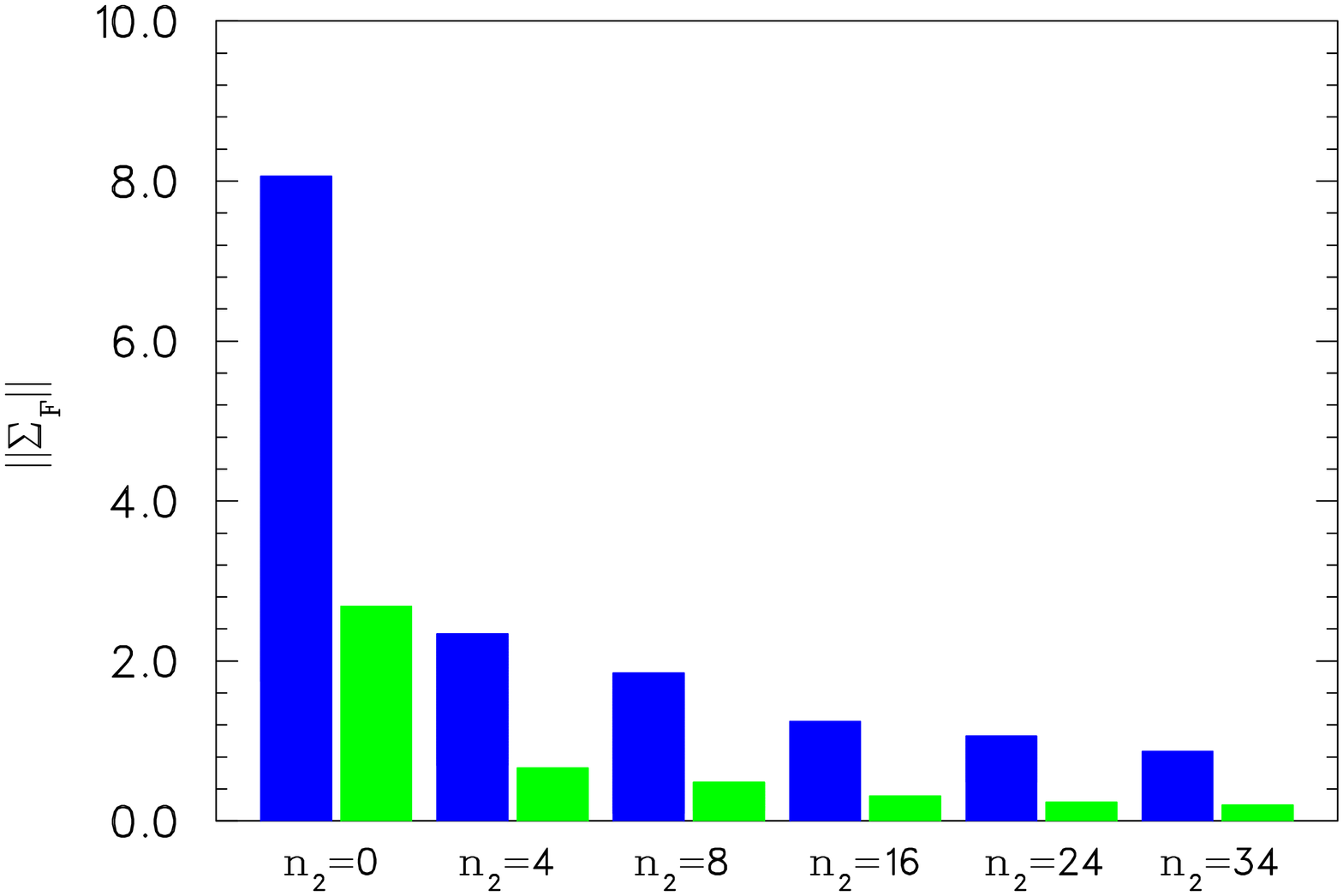}
\includegraphics[angle=90,height=0.28\textheight,width=0.45\textwidth]{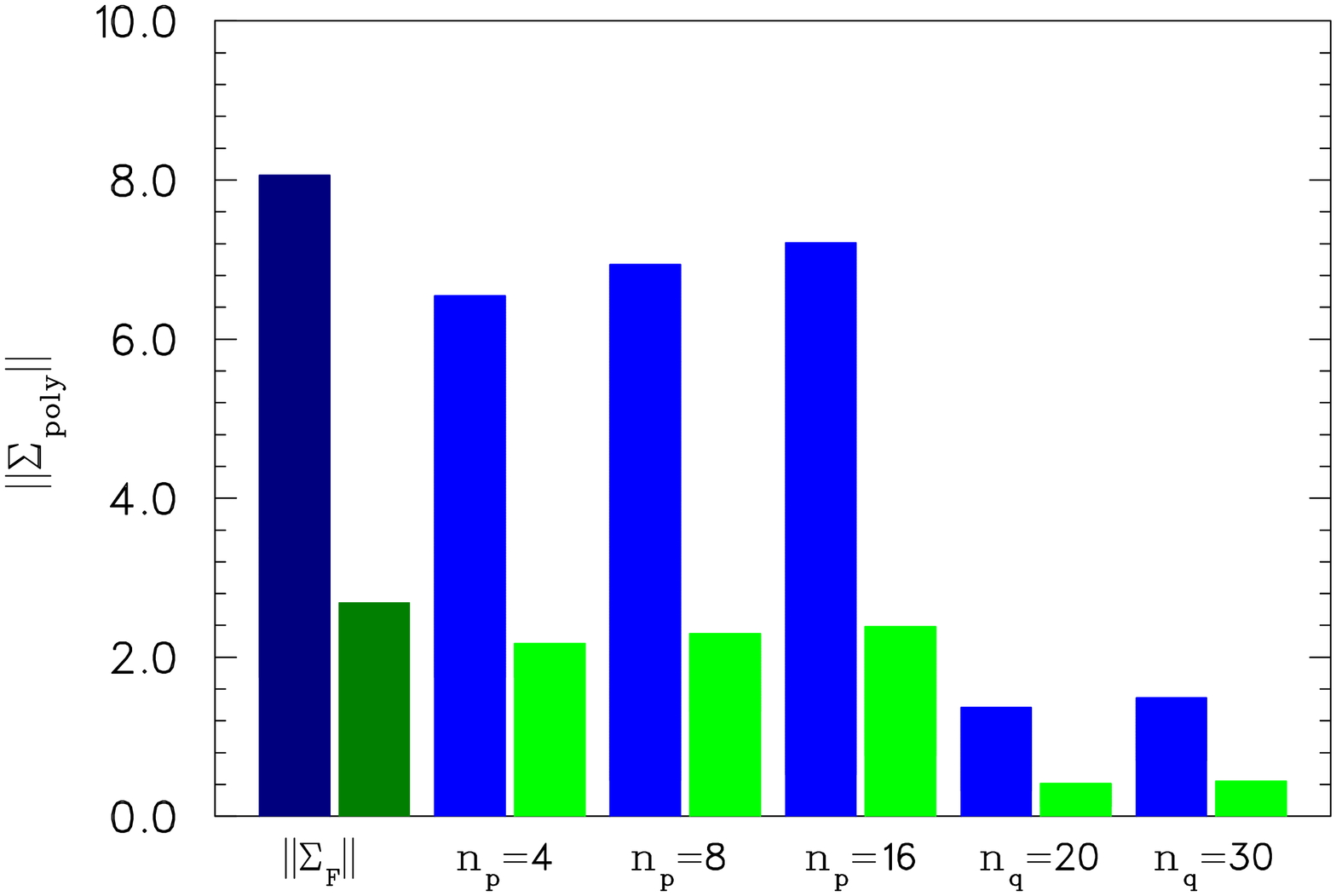}

\caption{\label{fig:force-pq2}The size of the force due to the pseudofermions $||\Sigma_F||$ (left) and that due to the polynomial term $||\Sigma_{poly}||$ (right), as a function of polynomial filter order. Results are shown for various $n_p,$ with two results for $n_q=20,30$ for which $n_p=4.$ The maximum and average force size is shown for $\kappa=0.15825.$ For comparison, the values for $||\Sigma_F||$ with $n_p=0,n_q=0$ are also shown as the leftmost darker coloured bars in the graph for $||\Sigma_{\rm poly}||$ (right). }
\end{figure*}

\begin{figure*}[!t]
\centering
\includegraphics[angle=90,height=0.28\textheight,width=0.45\textwidth]{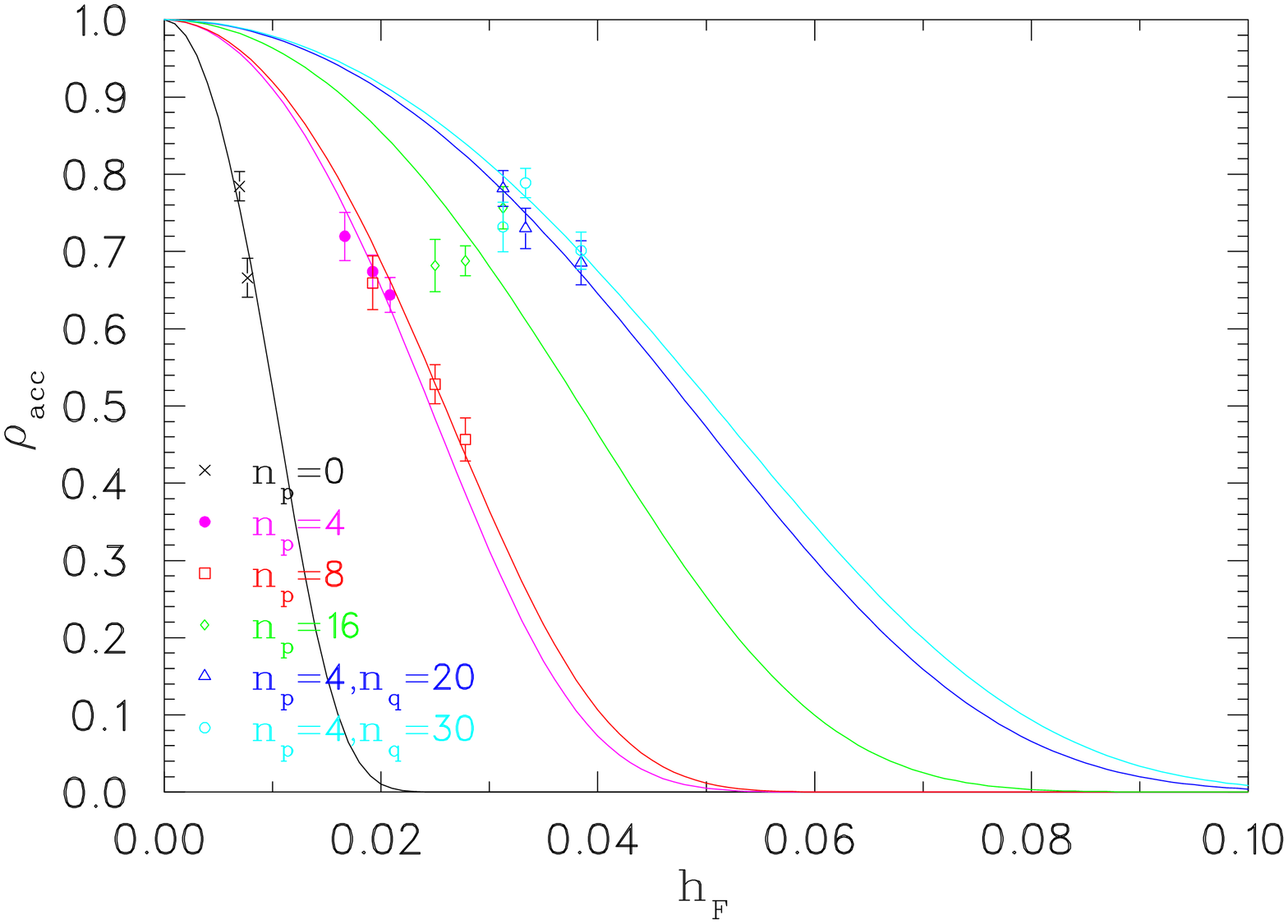}
\includegraphics[angle=90,height=0.28\textheight,width=0.45\textwidth]{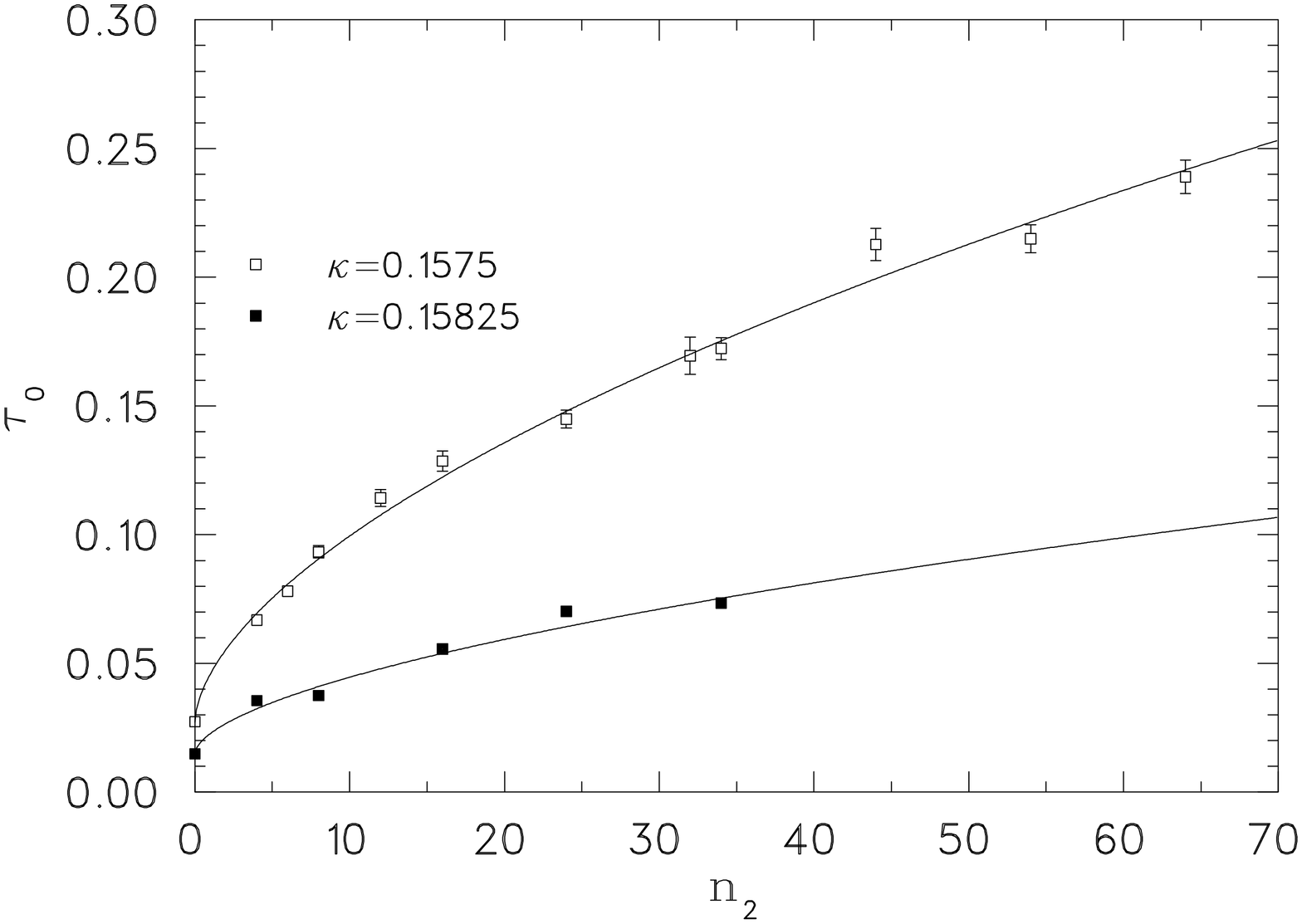}

\caption{\label{fig:accept-pq2}(Left) Trajectory acceptance rate as a function of pseudo-fermion step size for a dual polynomial filter of various order $n_p$ and $n_q,$ with $\kappa=0.1575.$ (Right) Fit of the characteristic scale $\tau_0$ as a function of polynomial filter order $n_2 = n_p+n_q.$ Combined results are shown for both $\kappa=0.1575$ and $\kappa=0.15825.$}
\end{figure*}

We repeat the comparison of force sizes, acceptance rates and characteristic
scales for our lighter quark mass at $\kappa=0.15825.$ Results are shown in
Figures~\ref{fig:force-pq2} and \ref{fig:accept-pq2} for a variety of polynomial filters. We see the same qualitative behaviour in these results as those for $\kappa=0.1575$ discussed above. In the right hand plot of Figure~\ref{fig:accept-pq2} fits for the characteristic scale for both masses using combined data for the single and dual polynomial filter results are shown. The values obtained for $a,b,c$ are given in Table~\ref{tab:taufit}.

\begin{table}[!t]
\centering
\begin{tabular}{ccccc}
$\kappa$ & $a$ & $b$ & $c$ & $\chi^2/\text{dof}$ \\
0.1575 & 0.0150(4) & 0.0778(5) & 0.581(2) & 2.12 \\
0.15825 & 0.0271(9) & 0.0187(9) & 0.586(1) & 15.8 \\
\end{tabular}
\caption{\label{tab:taufit} Values obtained for the fit parameters for $\tau_0 = a + b n_1^c.$}
\end{table}

\begin{figure*}[!t]
\centering
\includegraphics[angle=90,height=0.28\textheight,width=0.45\textwidth]{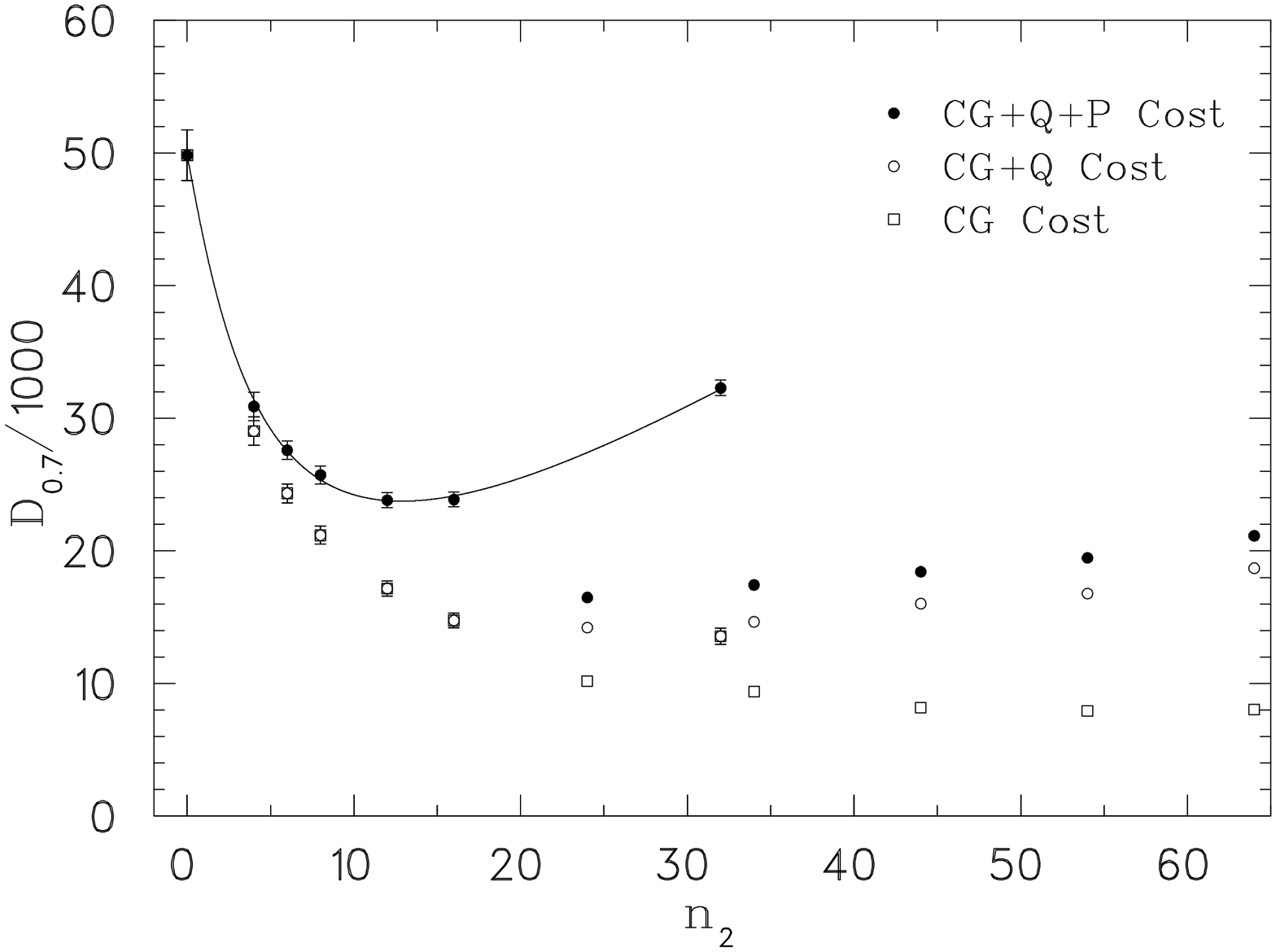}
\includegraphics[angle=90,height=0.28\textheight,width=0.45\textwidth]{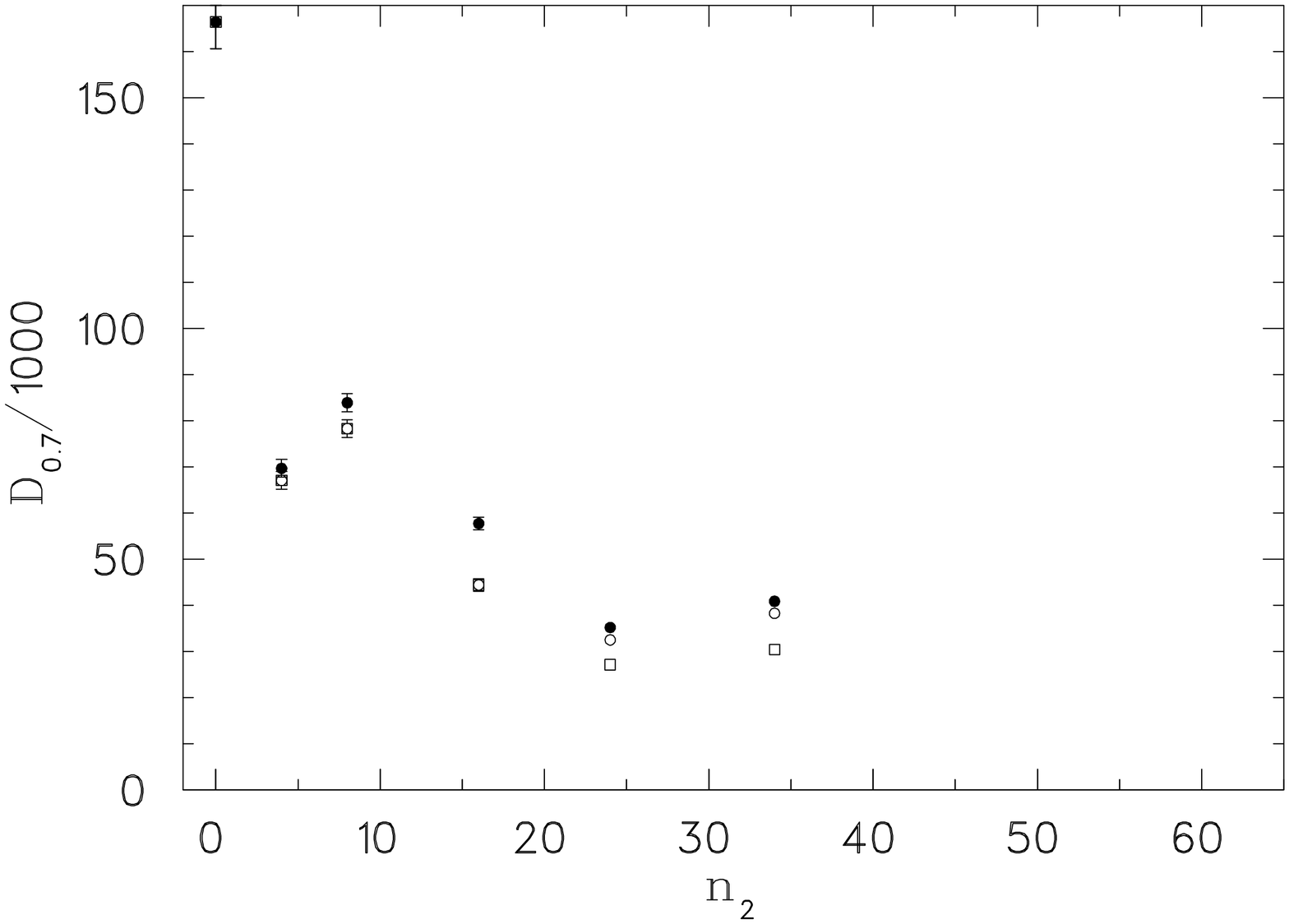}

\caption{\label{fig:cost} The cost $D_{0.7}$ as a function of polynomial filter size $n_2,$ for $\kappa=0.1575$ (Left) and $\kappa=0.15825$ (Right).}
\end{figure*}

In order to measure the relative numerical cost of the different choices for $n_p,n_q$ we calculate the total number of Dirac matrix applications needed per molecular dynamics trajectory at an acceptance rate of $0.7,$ and we denote this value by $D_{0.7}.$ Inverting Eq.~\ref{eq:rho} allows us to determine the pseudofermion step size $h_F$ that corresponds to $\rho_{\rm acc}=0.7.$ Keeping the polynomial step sizes approximately fixed at $N_P \approx 120, N_Q\approx 30,$ it is then a simple matter to calculate $D_{0.7}.$ Figure~\ref{fig:cost} shows $D_{0.7}$ as a function of polynomial order $n_2.$ We see that although there is a large initial reduction in cost, $D_{0.7}$ becomes nearly flat after about $n_2 = 4.$ The reason for this is that our polynomial of order $n_2$ reduces the condition number of the Dirac matrix by less than a factor of $(n_2+1)^{-1}.$ As each iteration of the conjugate gradient routine requires $n_2+1$ applications of the Dirac matrix, we can see that the cost of inverting the filtered Dirac matrix is increasing. This is offsetting the gain we obtain by having to invert less often. A potential improvement which could prove particularly useful for larger polynomials is to make use of a linear multi-shift inverter\cite{Frommer:1995ik,jegerlehner-multicg,Jegerlehner:1997rn} to evaluate the action of the inverse of $M\mathcal{P}.$

However, at large $n_2$ the force contribution from the fermions is very small indeed. This allows us to slacken the inversion target residual during the molecular dynamics integration. Slackening the residual from $10^{-7}$ to $10^{-4}$ significantly reduces $D_{0.7}.$ This is equivalent to working with an approximate Hamiltonian during the integration, and hence may put an upper bound less than one on the acceptance rate. This effect is observed for example if we try $r=10^{-3}.$

As an aside, we also performed some tests using a variant polynomial filter applied to the Wilson operator $M$ directly rather than $K=M^\dagger M.$ However, as one goes to light quark masses we find that the spectrum of $M$ (which is complex) intrudes upon the boundary of the elliptical region defined by the roots of the polynomial, making this alternative method unfeasible.

%% file: conclude.tex

The use of polynomials to separate different time scales in the molecular
dynamics integration step of the HMC algorithm applied to lattice QCD with 
dynamical quark fields was introduced. 
This shifts the most expensive force terms to the coarsest scale (and vice 
versa), allowing multiple time scale integration schemes to be effective. The 
procedure is extensible allowing not only the separation of UV and IR dynamics, 
but also the introduction of intermediate scales.  A novel generalisation of 
the leapfrog integrator was introduced which allows for far greater 
flexibility in choosing the time scale that one associates with the dynamics 
induced by a particular term in the action. The integrator is applicable to 
any simulation that makes use of multiple time scales.
 
Polynomial approximations to the inverse were shown to be successful UV filters 
for two flavour simulations. One possible improvement is to use a multi-shift 
solver to calculate the inversion of $M\mathcal{P}$. This should prove most 
advantageous when dealing with polynomials of larger order, which occur 
when using an intermediate filter. The use of a polynomial filter can be 
applied to single flavour simulations using a variety of implementations. A 
detailed description and comparison of these techniques are the subject of 
future work.

%% file: acknowledge.tex
We thank the Trinity Centre for High-Performance Computing (TCHPC) and the 
Australian Partnership for Advanced Computing (APAC) for 
providing access to the computational resources used for this work. The 
research was supported by Science Foundation Ireland under grants 
04/BR/P0266 and 07/RFP/PHYF168. 

%% file: appendix.tex
We perform an error analysis of our generalised integrator for a simple choice of stepsizes, following the procedure in \cite{sexton-weingarten}. Given a Hamiltonian $\mathcal{H}$ we can write the evolution operator for our system as $\exp{h\hat{\mathcal{H}}},$ with stepsize $h.$ Here we have defined $\hat{\mathcal{H}}$ as the linear operator on the vector space of functions $f$ on phase space $(p,q)$ defined by the Poisson bracket
\eqn{
\hat{\mathcal{H}}f = -\{ \mathcal{H}, f \}
= \sum_i \left( \frac{\del \mathcal{H}}{\del p_i}\frac{\del f}{\del q_i} - \frac{\del \mathcal{H}}{\del q_i}\frac{\del f}{\del p_i} \right).
}
If we write the Hamiltonian as 
\eqn{\mathcal{H}=T+S_1+S_2+S_3+S_4+\ldots}
then for each term in the Hamiltonian we can correspondingly define a
linear operator using the Poisson bracket relation above.

Proceeding with the error analysis, we make use of the Baker-Campbell-Hausdorff result,
\begin{multline}
e^{\lambda \hat{A}}e^{\lambda \hat{B}}e^{\lambda \hat{A}}=\exp\Big( \lambda(2\hat{A}+\hat{B}) + \\
\frac{\lambda^3}{6}([[\hat{A},\hat{B}],\hat{A}]+[[\hat{A},\hat{B}],\hat{B}])+O(\lambda^4)\Big)
\end{multline}
and apply this to the generalised leapfrog integrator in the simple case of $\mathcal{H}=T+S_1+S_2,$ where the time scale for each term in $\mathcal{H}$ corresponds to a number of integration steps $N_T=6,N_1=3$ and $N_2=2$ respectively. The integrator for this simplest non-trivial case can be written as
\eqn{ e^{\frac{h}{4}\hat{S}_2}e^{\frac{h}{6}\hat{S}_1}e^{\frac{h}{3}\hat{T}}e^{\frac{h}{3}\hat{S}_1}e^{\frac{h}{6}\hat{T}}e^{\frac{h}{2}\hat{S}_2}e^{\frac{h}{6}\hat{T}}e^{\frac{h}{3}\hat{S}_1}e^{\frac{h}{3}\hat{T}} e^{\frac{h}{6}\hat{S}_1}e^{\frac{h}{4}\hat{S}_2}.}
Repeated application of our BCH result allows us to deduce that the above expression can be written as
\begin{multline}
\exp\Big( h \hat{\mathcal{H}} + h^3\big(\frac{1}{48}[[\hat{S}_2,\hat{T}],\hat{T}]+\frac{1}{96}[[\hat{S}_2,\hat{T}],\hat{S}_2]+\\
\frac{1}{216}[[\hat{S}_1,\hat{T}],\hat{S}_1]+\frac{1}{108}[[\hat{S}_1,\hat{T}],\hat{T}]+\frac{1}{36}[[\hat{S}_1,\hat{T}],\hat{S}_2]\big)\Big)
\end{multline}
From this expression we can immediately see that the error in the generalised integrator relative to the leading term is $O(h^2),$ just as for the regular leapfrog.

If we examine the individual leapfrog integrators corresponding to
\eqn{\begin{matrix}H_1 = T + S_1,& H_2 = T + S_2\end{matrix}}
we obtain
\begin{multline}
e^{\frac{h}{6}\hat{S}_1}e^{\frac{h}{3}\hat{T}}e^{\frac{h}{3}\hat{S}_1}e^{\frac{h}{3}\hat{T}}e^{\frac{h}{3}\hat{S}_1}e^{\frac{h}{3}\hat{T}}e^{\frac{h}{6}\hat{S}_1}\\=\exp\Big( h \hat{H}_1 +
h^3(\frac{1}{108}[[\hat{S}_1,\hat{T}],\hat{T}]+ \frac{1}{216}[[\hat{S}_1,\hat{T}],\hat{S}_1])\Big),
\end{multline}
and
\begin{multline}
e^{\frac{h}{4}\hat{S}_2}e^{\frac{h}{2}\hat{T}}e^{\frac{h}{2}\hat{S}_2}e^{\frac{h}{2}\hat{T}}e^{\frac{h}{4}\hat{S}_2}=\exp\Big( h \hat{H}_2 + \\
h^3(\frac{1}{48}[[\hat{S}_2,\hat{T}],\hat{T}]+ \frac{1}{96}[[\hat{S}_2,\hat{T}],\hat{S}_2])\Big)
\end{multline}

Hence we see that the only difference between the individual integrators and our generalised integrator is the cross term $[[\hat{S}_1,\hat{T}],\hat{S}_2].$ The algorithm is identical to a standard nested leapfrog in the case where $N_i|N_{i-1}.$